\RequirePackage{graphicx}
\documentclass[aps,pra,floatfix,reprint,longbibliography]{revtex4-1}
\usepackage{enumitem}
\usepackage{mathtools}
\usepackage{xcolor}
\usepackage{float}
\usepackage{tikz}
\usetikzlibrary{quantikz}
\usepackage{graphicx}
\usepackage{natbib}
\usepackage{dcolumn}
\usepackage{bm}
\usepackage{mathtools}
\usepackage{braket}
\usepackage{algorithm}
\usepackage{algpseudocode}
\usepackage[utf8x]{inputenc}
\usepackage{amsmath}
\usepackage{multirow}
\usepackage{amsfonts}
\usepackage{amssymb}
\usepackage{rotating} 

\usepackage{amsthm}
\date{\today}
\newcommand{\Gam}[4]{\hat{a}^\dagger_{#1} \hat{a}^\dagger_{#2} \hat{a}^{}_{#3} \hat{a}^{}_{#4}}

\begin{document}
\title{Many-Body Excited States with a Contracted Quantum Eigensolver}
\author{Scott E. Smart, Davis M. Welakuh, Prineha Narang}
\affiliation{College of Letters and Science, Physical Sciences Division, University of California, Los Angeles, USA}

\begin{abstract}
Calculating ground and excited states is an exciting prospect for near-term quantum computing applications, and accurate and efficient algorithms are needed to assess viable directions. We develop an excited state approach based on the contracted quantum eigensolver (ES-CQE), which iteratively attempts to find a solution to a contraction of the Schr{\"o}dinger equation projected onto a subspace, and does not require a priori information on the system. We focus on the anti-Hermitian portion of the equation, leading to a two-body unitary ansatz. We investigate the role of symmetries, initial states, constraints, and overall performance within the context of the model rectangular ${\rm H}_4$ system. We show the ES-CQE achieves near-exact accuracy across the majority of states, covering regions of strong and weak electron correlation, while also elucidating challenging instances for two-body unitary ansatz. 
\end{abstract}
\maketitle


\section{Introduction}
Quantum simulation offers a unique pathway towards studying the many-body problem. The success of variational and adiabatic approaches have led to focuses on ground state approaches \cite{mcardle2020rmp,tilly2020,Head-Marsden2020,bauer2020}, but for many physically relevant phenomena a robust characterization of excited states is needed, corresponding to the obtaining arbitrary eigenstates of the Hamiltonian \cite{Abrams1999}. Spectroscopic applications, photochemical and biomolecular processes \cite{cheng2009arpc, gonzalez2012c}, light-matter interactions \cite{ruggenthaler2018nrc,liebenthal2022}, open-systems simulation\cite{Head-Marsden2020prr} and non-adiabatic configurations \cite{matsika2021cr,domcke2004,butler1998} all require accurate information on the spectra of states. Numerous classical methods exist for different systems within physics and chemistry \cite{dreuw2005cra, lischka2018cr}, and in general these are more complex than the ground state problem. While quantum algorithms for ground and excited state both utilize direct access to a quantum state, excited state approaches often benefit from the capacity to store and manipulate several states simultaneously. 


Two broad classes of excited state methods exist on quantum computers, namely reference-based expansions or state-based methods. The reference-based approaches generate a set of trial states, either from a single wavefunction \cite{McClean2017,Colless2018,yoshioka2022prlb}, or multiple wavefunctions which can be generated systematically via time evolution, imaginary time evolution, state-averaged minimization, or other means \cite{Huggins2020,baek2022,baker2021a,Motta2019,parrish2019prl,parrish2019,klymko2022pq}. From these, one generally solves a generalized eigenvalue equation. Alternative techniques include equation of motion \cite{ollitrault2020prr,asthana2023cs} or quantum linear response approaches \cite{chiew2023a}. These can yield entire sets of properties, such as excitation energies, dipole moments, response properties, etc., but do not always provide verifiable information on the states.  

State-based methods focus on obtaining eigenstate estimates by iteratively solving for states on a quantum computer. A well known approach is the variational quantum deflation (VQD) technique, which iteratively remove the influence of a state with a variational quantum eigensolver-based (VQE) optimization \cite{higgott2018,Lee2018,kottmann2021cs,gocho2023ncm,jouzdani2019b}. The subspace-search \cite{nakanishi2019prr} and orthogonal state-reduction VQE \cite{xie2022jctc} provide expanded frameworks for ancilla-based optimizations, while constrained-VQE adds additional constraints \cite{Ryabinkin2019}. Other techniques have investigated improving the standard unitary coupled cluster singles and doubles (UCCSD) ansatz, such as ADAPT-VQE expansions \cite{yordanov2022pra,chan2021pccpa}, iterative generalized pair-cluster ansatz \cite{Lee2018}, or symmetry-based restrictions on the ansatz \cite{gocho2023ncm,lyu2023q}. While these methods are robust, and iteratively can improve accuracy, with the exception of the ADAPT-VQE, they offer limited information on the reliability of the wavefunction. Other techniques include the witness-assisted variational scheme \cite{santagati2018}, or the variance-VQE \cite{zhang2022cpb}, which attempt to minimize the variance of the state, can locate eigenstates. State-based and reference based approaches have both been demonstrated on quantum devices \cite{ollitrault2020prr,motta2023cs,yeter-aydeniz2020nqi}, and other techniques beyond these (including machine-learning based) exist as well \cite{kawai2020mlst,sureshbabu2021}.

The contracted quantum eigensolver (CQE )\cite{smart2021prl,smart2021pra,smart2022jctc} is an alternative framework to VQE which attempts to iteratively reduce the residual of a contracted equation via an exponential ansatz. The residuals relate to gradients in the exponential space, which can be measured efficiently on a quantum computer \cite{smart2021prl}, and also provide a stationarity condition on the wavefunction \cite{nakatsuji1976pra,mazziotti2004pra}. Classically, these methods relate to techniques within reduced density matrix (RDM) approaches, most notably methods based on solving the contracted Schr{\"o}dinger equation \cite{nakatsuji1976pra,nakatsuji2004prl,mazziotti2006prl,mazziotti2007pr-amop}. For the quantum algorithms, the primary cost in the near-term era is the length of the ansatz, though approximate update schemes and compilation strategies can be used to reduce these costs, and are an ongoing area of research \cite{smart2022jctc,rubin2022jctc}. While these methods have been successfully applied for ground states, excited state based approaches have limited classical analogues, as these can require obtaining transition density matrices (TDM) \cite{sand2015jcp,booth2014mp,etienne2015jcp,mazziotti1999pra,boyn2021jcp}.

The current work introduces an excited state contracted quantum eigensolver (ES-CQE). We explore a projection of the Schr{\"o}dinger equation (which removes the need for prior information) as well as deflation based techniques for constraining the problem. Using the ES-CQE based on the anti-Hermitian form, we find that we are able to achieve essentially exact solutions compared to the full configuration interaction (FCI) method, except in cases where the symmetries of the system are non-trivial. We demonstrate the algorithm with the rectangular ${\rm H}_4$ molecular, which has a diverse array of characteristics in the ground and excited state regime related to electron correlation. We also provide comparisons of the optimization strategy as and the accuracy compared to other common methods. 

\section{Excited State Contracted Quantum Eigensolver}
We first review the many-body eigenstate problem, and then review projections of the Hamiltonian and Schr{\"o}dinger equation and the resulting variational formulations. We then introduce an exact form based on a contraction of the projected Schr{\"o}dinger equation and derive the corresponding quantum algorithm, and end with some practical modifications of the algorithm. 

In the many-body problem we often are attempting to find a solution $|\psi_\rangle$ to the electronic Hamiltonian, which comprises the electrons centered at a given nuclear configuration \cite{helgaker2000met}:
\begin{equation}
 \hat{H} = \sum_{pq} {}^1 K^{p}_{q} a^\dagger_p a^{}_{q} + \frac{1}{2}\sum_{pqrs} {}^2 V^{pr}_{qs} \Gam{p}{r}{s}{q}.
\end{equation}
Here, $p,q,r,s$ are spin orbital indices in a given orbital representation, $a^\dagger_{p}$ and $a^{}_{p}$ are fermionic creation and annihilation operators acting on an orbital $p$, ${}^2 V^{pq}_{rs}$ are the two-electron integrals and ${}^1K^p_q$ contains the electron-nuclear interaction and the electron kinetic energy. The ground state solution is the lowest energy eigenstate, while other eigenstates are referred to as excited states. 

While the algorithm here focuses on molecular systems with only two-body molecular interactions, the technique can be generalized to other systems. For the dispersion relation to be satisfied, we simply require that the set of contractions (or residuals) spans the same space as the observable. 

\subsection{Subspace Projections of the Schr{\"o}dinger Equation }
In the variational quantum deflation scheme \cite{higgott2018,Lee2019}, the Hamiltonian of a given system is iteratively deflated by biasing the influence of the target states \cite{mackey}.
In principal component analysis, the deflation procedures assigns projected eigenvectors a null eigenvalue \cite{mackey}. On a quantum computer, these projected states can still infiltrate the optimization, and so in the VQD approach a constraint based on the energy gap (or worst case, the Hamiltonian norm) is added. 

Given a set of projected states denoted by $\mathcal{A}$, and a projection operator $\hat{P}_\mathcal{A}$ projecting onto the complement of $\mathcal{A}$, 
\begin{equation}
    \hat{P}_\mathcal{A} = \mathcal{I} - \sum_\alpha |\alpha \rangle \langle \alpha |, 
\end{equation}
the deflated energy expression is:
\begin{equation}\label{vqd}
    E = \langle \psi |\hat{H} \hat{P}_\mathcal{A} |\psi \rangle .
\end{equation}
Applying the variational principle we find that:
\begin{equation}
    \delta E = \langle \psi | \hat{H} | \delta \psi \rangle - \sum_\alpha E_\alpha \langle \psi | \alpha \rangle \langle \alpha | \delta  \psi \rangle + h.c. 
\end{equation}
where h.c. denotes the Hermitian conjugate terms. 

We can instead apply the projection to the entire Schr{\"o}dinger equation (SE), effectively solving it in a subspace. We denote the projected Schr{\"o}dinger equation (PSE) as:
\begin{equation}
    \hat{P}_\mathcal{A} (\hat{H} - E) |\psi \rangle = 0.
\end{equation}
where we only specify that $\hat{P}|\psi\rangle \neq |0\rangle $, i.e. a portion of $|\psi\rangle$ is in the image of $\hat{P}$. 

It is easily verified that the PSE is equivalent to the SE when the projected states are eigenstates. However, any combination of an eigenstate $|\phi\rangle$ with another eigenstate state $|\chi\rangle \in \mathcal{A}$ also satisfies the PSE. Thus, a trial ansatz on a quantum computer is not guaranteed to actually be orthogonal to $\mathcal{A}$, which can lead to numerical problems. Practically we address this by adding a constraint to the state overlaps:
\begin{equation}
    c_\alpha(\psi) = 1 - \langle \psi | \alpha \rangle \langle \alpha |\psi \rangle   .
\end{equation}
When these constraints are strongly violated, we can either increase measurement results, or reject the normalization and use the form in Eq.~\eqref{vqd} with a high penalty until we get out of a problematic region. 

Looking at variations of the energy w.r.t. $|\psi\rangle$ in the PSE, we find that:
\begin{align}\label{eq:vpse}
\delta E =& \frac{\langle \psi | \hat{P}_\mathcal{A} \hat{H} | \delta \psi \rangle - E \langle \psi | \hat{P}_\mathcal{A} | \delta \psi \rangle }{\langle \psi | \hat{P}_\mathcal{A} | \psi \rangle} + h.c. \\ 
    =& \frac{\langle  \psi | \hat{H} - E | \psi \delta \rangle + \sum_\alpha (E - E_\alpha)\langle \psi | \alpha \rangle \langle \alpha |  \delta \psi \rangle }{1- \sum_\alpha |\langle \psi | \alpha \rangle |^2} + h.c.
\end{align}
The relation above applies broadly to variational algorithms, including the VQE and CQE approaches. Here, the role of $\beta_i$ in VQD can be compared with the $E - E_i$ in the energy evaluation. With any additional constraint, we naturally generate the energy inequality. While energy gaps are still implicitly present in the optimization, the optimization is carried out \emph{ab initio}, with no explicit knowledge of the system required. 
 
\subsection{Contracting Projected Forms of the Schr{\"o}dinger Equation}
We contract the PSE by introducing a set of two-electron test functions \cite{mazziotti1998pra}. Similar to the contraction of the Schr{\"o}dinger equation, we denote this as the contracted projected Schrodinger equation (PCSE)\cite{nakatsuji1976pra,mazziotti1998pra}.

The CPSE is defined as:
\begin{align}
    \langle \psi |\hat{P}_\mathcal{A}(\hat{H} - E)  \Gam{i}{k}{l}{j} |\psi \rangle N^{-1} = {}^2 R^{ik}_{jl}.
\end{align}
where $i,j,k,l$ are spin orbital indices, and $N$ represents the norm:
\begin{equation}
    N = 1 - \sum_{\alpha} \langle \psi | \alpha \rangle \langle \alpha | \psi \rangle.
\end{equation} 
The elements ${}^2 R^{ik}_{jl}$ are denoted as the residuals of the contracted equation for a given set of indices. When all residuals are zero, we obtain a solution of the contracted equation and the projected Schr{\"o}dinger equation. 

This condition can also be seen by inserting the identity $\Gam{i}{k}{l}{j} = \frac{d}{d {}^2 J^{ik}_{jl}} e^{\epsilon \hat{J}}|_{\epsilon=0}$, where $\hat{J}$ is a generic two-body operator:
\begin{equation}
    {}^2 \hat{J} = \sum_{ikjl} {}^2 J^{ik}_{jl}\Gam{i}{k}{l}{j} 
\end{equation}
into Eq.~\eqref{eq:vpse} as the variation of $\psi$, relating the PCSE to the infinitesimal expansion of the projected energy around a two-body operator:
\begin{equation}  {}^2 R^{ik}_{jl} =  \frac{d E}{d ~ {}^2 J^{ik}_{jl}} \frac{dE}{d\epsilon} \frac{\langle \psi |e^{\epsilon J^\dagger} \hat{P}_\mathcal{A} \hat{H}  e^{\epsilon J}| \psi \rangle }{\langle \psi |e^{J^\dagger} \hat{P}_\mathcal{A} e^{J}| \psi \rangle}.
\end{equation}

Explicitly, the residual equation has the form:
\begin{align}
\begin{split}
{}^2 R^{ik}_{jl} N  =& \langle \psi | \hat{H} \Gam{i}{k}{l}{j}| \psi \rangle - {}^2 D^{ik}_{jl} E \\ &+ \sum_\alpha (E-E_\alpha) \langle \psi |\Gam{i}{k}{l}{j} |\alpha \rangle \langle \alpha | \psi \rangle 
\end{split}
\end{align}
where the final term contains the 2-electron transition density matrix (2-TDM) times an overlap. This equation is a necessary and sufficient condition for $|\psi\rangle$ to describe an eigenstate in the projected space. We can take the anti-Hermitian contribution of this, yielding:
\begin{align}
\begin{split}  {}^2 A^{ik}_{jl} N &= \langle \psi | [\hat{H},\Gam{i}{k}{l}{j} ] | \psi  \rangle \\ 
 &+ \sum_{\alpha}( E -E_\alpha ) \big(\langle \psi | \Gam{i}{k}{l}{j} |\alpha \rangle  \langle \alpha  |\psi \rangle \\ 
 &~~-  \langle \alpha | \Gam{i}{k}{l}{j} |\psi \rangle  \langle \psi |  \alpha \rangle  \big) . \end{split}
\end{align}
In the present work we denote this as the ACPSE, the anti-Hermitian contracted projected Schr{\"o}dinger equation. These anti-Hermitian residuals ensure that $\exp {}^2 J $ is a unitary operator, and can be considered a restriction of the ACSE condition \cite{kutzelnigg1979cpl,mazziotti2007pr-amop} to a projected space. 

While classically the CSE and ACSE equations can be solved using the 4- and 3-electron reduced density matrices respectively, the CPSE and ACPSE contain a 2-TDM. TDMs are generally non-Hermitian and are defined between two states usually with a known, simple reference\cite{booth2014mp,etienne2015jcp,mazziotti1999pra}. To the best of the authors knowledge, there is currently no low cost technique for generating these matrices for two generic states without reference to the wavefunctions. As a result this algorithm is particularly well suited for quantum computers\cite{smart2021pra}.
 
 If instead of the CPSE, we want the deflated form, we can simply contract the projected Hamiltonian. The resulting energy is given as:
\begin{equation}
E(\epsilon, \hat{A}) = \langle \psi |e^{-\epsilon A} \hat{H} \mathcal{P}  e^{\epsilon A} |\psi \rangle ,
\end{equation}
with residuals:
\begin{align}
\begin{split}
A^{ik}_{jl} = \frac{d}{d\epsilon}\frac{dE}{d A^{ik}_{jl}} &= \langle \psi | [\hat{H},\Gam{i}{k}{l}{j}]| \psi \rangle \\ 
   & - \sum_\alpha E_\alpha \langle \psi | \Gam{i}{k}{l}{j} | \alpha \rangle \langle \alpha | \psi \rangle \\
   & + \sum_\alpha E_\alpha \langle \psi | \alpha \rangle \langle \alpha | \Gam{i}{k}{l}{j}  | \psi \rangle.
\end{split}
\end{align}
These forms were used in context of operator selection in the ADAPT-VQD ansatz at each macro iteration\cite{chan2021pccpa}. Without an additional constraint, this has the effect of needing to introduce an energy dependent constraint onto the equation to aid convergence. 

\begin{figure*}
    \centering
    \includegraphics[scale=0.27]{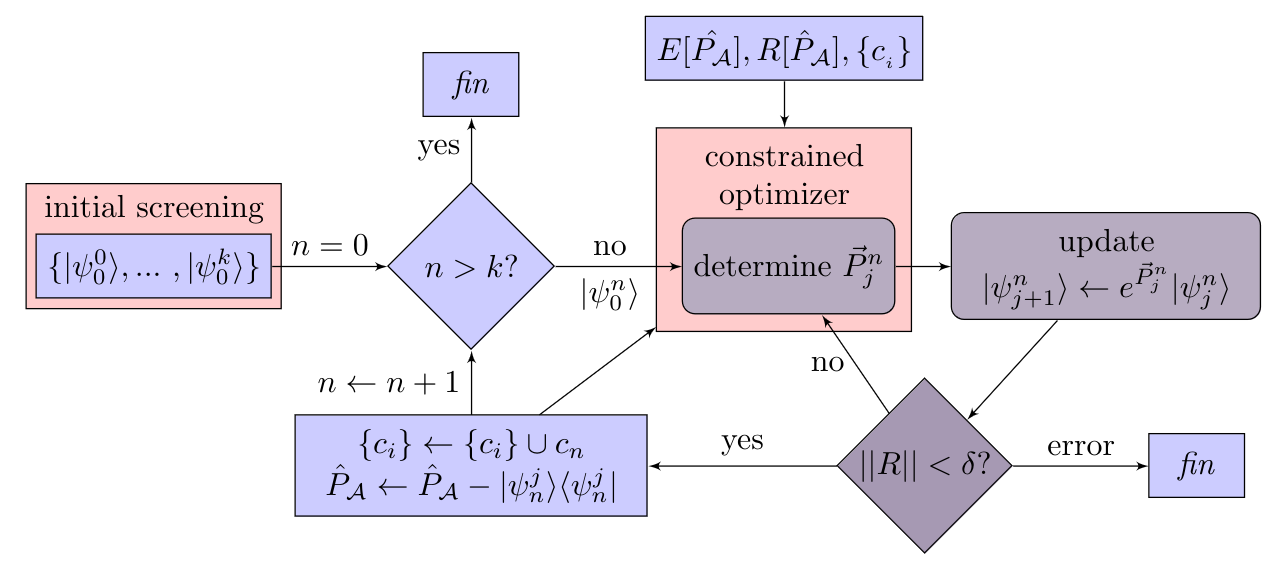}
    \caption{Overview of excited state CQE approach. Given an initial set of $k-$states, and a method of projection, we perform $k$ CQE procedures, iteratively updating the constraints and projection as each state is found. In each CQE run, we iteratively determine the search direction (using a classical optimizer) and update the wavefunction.}
    \label{fig:my_label}
\end{figure*}
 
\subsection{Excited State CQE}
The excited state CQE algorithm is summarized in Figure 1. Given a set of $k$ initial states, we perform $k-$CQE runs, where the projected energy and residuals of the contracted equation serve as the function and gradient of a locally parameterized optimizer. The constrained problem is cast as an unconstrained one through one of several techniques, and the procedure then follows in a similar manner to the locally-parameterized CQE (using an optimizer such as BFGS or l-BFGS)\cite{smart2022jctc}.

At each step of the optimizer, we compute a search-direction, performing the optimization in the tangent space of the exponential. The result is an iterative wavefunction structure:
\begin{equation}
    |\psi_{n+1}\rangle = e^{\hat{P}_n} |\psi_n \rangle 
\end{equation}
where $\hat{P}$ is a two-body anti-Hermitian operator. Practically, we implement the terms in $\hat{P}$ linearly via a first-order trotterization with Pauli gadgets, as trotterization errors are largely negligible through the iterative nature of the algorithm and often smaller step sizes. Once a state is converged (i.e. the residual norm is below a threshold), we update the constraints of the problem, possibly terminating or adjusting constraints depending on the form of the optimizer. 

The projected energy can be obtained with the energy of the state subtracting the projected energies. The residuals require two components. The first is the ACSE, which can be recovered from a higher-order measurement on the quantum computer, or by applying infinitesimal based approaches scaling with the 2-RDM \cite{smart2021prl,schlimgen2022prr}. The second terms require the overlap and 2-TDMs between previous states, which we discuss below. 

Previous work \cite{smart2022jctc} discusses compilation strategies for reducing circuit length, particularly by approximate addition schemes so that terms in $\hat{P}_n$ are included as updates to previous terms, which can be substituted into the ES-CQE easily. In the present work we attempt to assess the accuracy, limitations and viability of the CQE for excited states, and less focus on the gate count of the algorithm. 

\subsubsection{Overlap and Transition Circuits}

The overlap and TDMs can be obtained in a straightforward manner on a quantum computer. The overlap circuit is performed with a single expectation value for each state, and the TDMs have the same measurement complexity as RDM measurements. For the $k-$th state this means calculating $k-1$ overlaps and $k-1$ TDMs. Practically there are several approaches we can take to reduce this, though the effective scaling is similar, leading to a quadratic cost with the current number of states $k$. 

Given the application of $N$ Pauli gadgets with an average CNOT cost of $M$ (for a total circuit cost $NM$),  we apply the overlap and TDM circuits with ideally $N(2M+1)$ CNOTs using one ancilla qubit. To do this, we first assume that we can decompose our set of operators as Pauli gadgets. Then, for each term we write the exponential of a Pauli string $\hat{O}$ as:
\begin{align}
    |0\rangle \langle 0| \otimes e^{i\theta \hat{O}} + |1 \rangle \langle 1 | &= e^{i{\theta |0\rangle\langle 0 | \otimes \hat{O} }} \\
    &= e^{i\frac{\theta}{2}\hat{I}\otimes \hat{O}} e^{i\frac{\theta}{2}\hat{Z}\otimes \hat{O}}. 
\end{align}
which conditions the Pauli on the $|0\rangle$ ancilla state. If we then denote the series of Pauli gadgets required to simulate state $k$ as $U_i^k$, we can write the overlap between two states as:
\begin{align}
    |\Psi \rangle &= \frac{1}{\sqrt{2}}(|0\rangle |\psi^k \rangle + |1\rangle|\psi^j \rangle ) \\
    \begin{split}
        &=\frac{1}{\sqrt{2}}(\prod_{i=0}( |0\rangle \langle 0 | \otimes  U^{k}_i) |0\rangle|\psi_0^k\rangle   \\ &~~~~+\prod_{i=0} (|1\rangle \langle 1| \otimes U^{j}_i)| 1\rangle|\psi_0^j\rangle ). 
    \end{split}
\end{align} 
As the terms at different iterations $U_i^k$ and $U_{i'}^k$ do not necessarily commute, the $I$ and $Z$ Pauli terms are implemented at each step. A measurement in the ancilla space of the Pauli $X$ or $Y$ gates provides the real and imaginary overlap components respectively. The transition 2-RDM measurements involve the product of this with the 2-RDM operator: $\hat{X} \otimes \Gam{i}{k}{l}{j}$ (or $\hat{Y}$). Naively this scheme quadruples the circuit cost, requiring $I$ and $Z$ combinations for both the target and constrained state. However, because the Pauli gadgets at each iteration are the same except on the ancilla, the circuit length can be shortened through compilation, ideally requiring only $O(1)$ extra CNOT gates. This effectively requires just the CNOT cost for the two states.  

\subsubsection{Initial Guess Procedure, CQE+}
Despite general confidence in the ground state as the global minima, in most variational procedures we depend on the initial state having sufficient overlap with the ground state. This problem is accentuated in excited state approaches, and so reliably obtaining relative orderings is a key feature of an excited state algorithm. This is also necessary to avoid large overlaps between the target state and the set of constraints, and can affect the length of the optimization. Towards this we tested several heuristics, with the most straightforward being to check a large number of simple states, and then sort the states according to their energies. 

For initial states we use single-reference Slater-determinant wave functions, as well as configuration state functions (CSFs, see Appendix for implementation details). Slater determinants energies are obtained very easily from the mean field states, and configuration state functions for low-spin configurations energies can also be evaluated quickly. 

While this initial ordering provides a more systematic way of choosing initial guesses, practical calculations show this is not always sufficient. In addition to this initial sorting we use a pre-screening procedure, which we refer to as CQE$+$. Given a selection of $k-$states, we perform a low-iteration and low-threshold CQE targeting $2*k$ states, and then reorder states accordingly. The rationale is that the largest energy gains are generally seen in the first few iterations, and so we allow the states to reorder themselves roughly away from the mean-field solution space. Because energy differences are related to the residual norms, we can also use that information as a cut-off as well. 

\subsubsection{Constrained Optimization}

While the PSE is not a constrained optimization, to ensure that we maximize the norm of the projected wavefunction, it is necessary to constrain the overlap condition, which modifies both the function and gradient evaluations. Instead of further expanding the ancilla space to insert orthogonality conditions from the trial problem \cite{nakanishi2019prra}, we simply add a small constraint in our optimization procedure. Importantly, the added constraint relates to the norms of the overlaps, and not the energy gaps. We tested basic penalty and Lagrangian multiplier forms, as well as the augmented Lagrangian approach \cite{Robinson2006}, noting that more specific techniques were recently introduced in the context of the deflation-based technique \cite{kanno2023}. 

The CPSE itself provides a necessary and sufficient condition for $|\psi\rangle$ to be an eigenstate within the projected space. For a noisy projection or a projection of non-eigenstates, these eigenstates may not correspond to solutions of Schr{\"o}dinger equation. This is not as much of a concern for low-noise calculations, but for noisy results, or results dominated by $k>2$ order excitations, this could pose a problem. To rectify this, we also pursued a relaxation procedure, whereby the constraints can be relaxed as we approach a solution. Properly this involves both a relaxation of the imposed constraints as well as a relaxation of the PCSE to the CSE. In practice, we found that when the state was properly orthogonal, the obtained results were highly accurate (i.e., true eigenstates), and so we leave these investigations for future work with noisy systems. 

\section{Results and Discussion}

\begin{figure}
    \centering
    \includegraphics[scale=0.83]{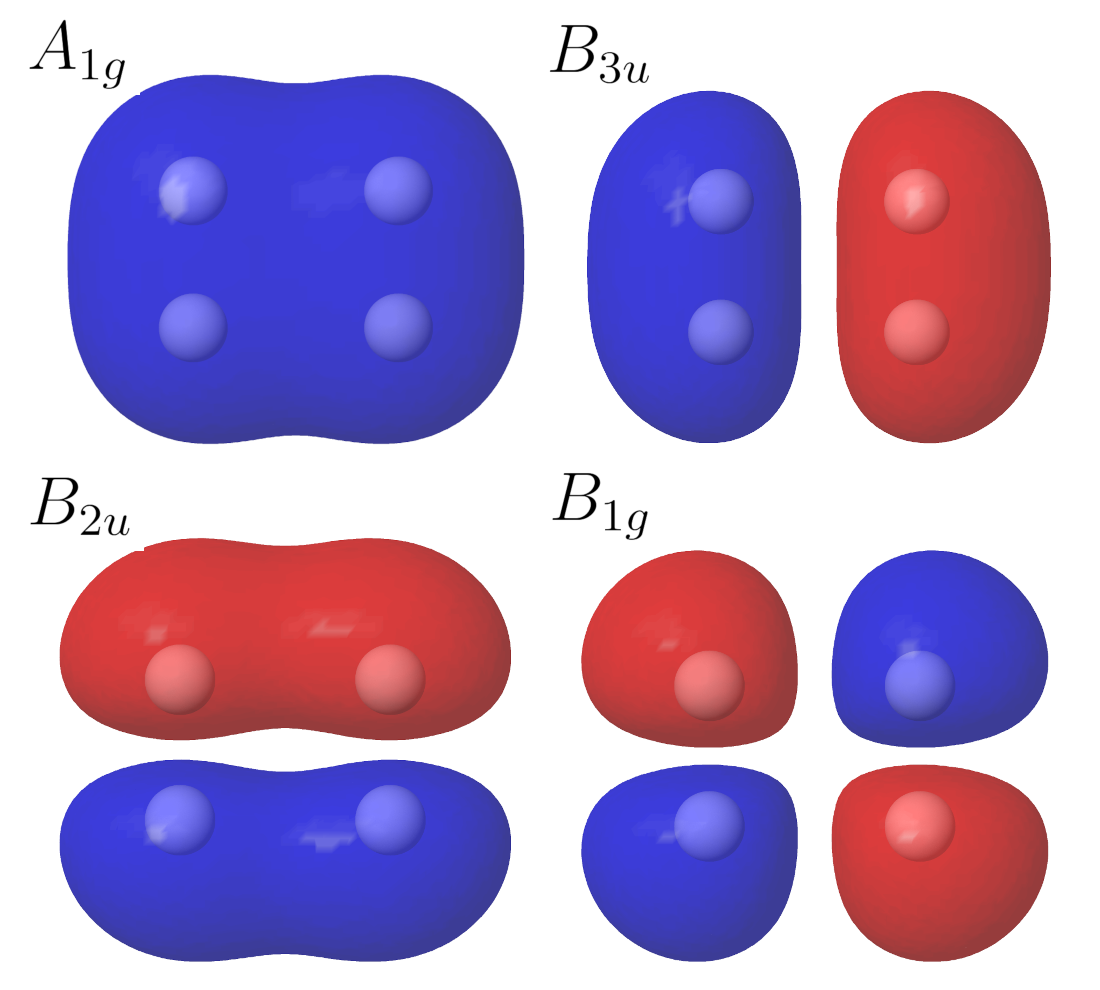}
    \caption{Molecular orbitals of rectangular ${\rm H}_4$ with side lengths 1.0 \AA~and 1.5 \AA, $D_{2h}$ point group. Also given are the molecular orbital irreducible representations. Energy increases from $A_{1g}$, $B_{3g}$, $B_{2u}$, to $B_{1u}$.}
    \label{fig:h4}
\end{figure}

The main objective in this work was to determine the efficacy of the CQE-based approach in a broad scheme. Because the ACSE represents a 2-particle condition, for any 2-electron systems, the ACPSE is an exact condition. Constant $S_z$, minimal basis $3-$electron systems are also restricted to double excitations, and so the ACPSE yields exact results. With four electron systems however, we begin to probe limitations in the two-electron ansatz. Thus, most of the work here covers the four-electron rectangular configuration of ${\rm H}_4$, seen in Figure~\ref{fig:h4}. Linear ${\rm H}_4$, and several molecular dimers exhibit well behaved excited state behaviors in their stretching and dissociation, while rectangular ${\rm H}_4$ is known to exhibit strongly correlated effects, particularly near the square geometry\cite{sand2015jcp,Finley1995,Romero2017}.

\subsection{Initial Comparison and Accuracy of ES-CQE}
\begin{figure}
    \centering
    \includegraphics[scale=0.58]{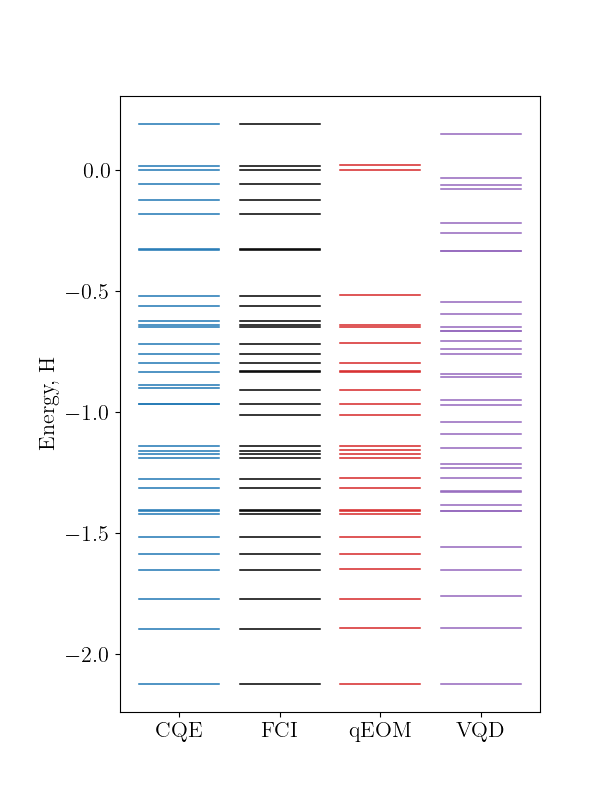}
    \caption{Comparison of full-state spectra obtained with the CQE-based approach, VQD given a UCCSD ansatz, and the qEOM approach for rectangular ${\rm H}_4$ at 1.0, 1.5~\AA ~side lengths. The qEOM is limited by the excitation manifold, but for the most part does well in estimating the relative energies. The VQD and CQE procedures are able to recover each state, though the VQD has significant errors throughout, and the CQE the entire spectrum with three outlier states.}
    \label{fig:2_method_compare}
\end{figure}

We first look at the excited state spectra obtained at 1.0 \AA~ and 1.5 \AA~ side lengths for ${\rm H}_4$ with the ES-CQE approach, the VQD approach, the quantum equation-of-motion (qEOM) approach (a reference based-expansion), and the FCI solution in Figure \eqref{fig:2_method_compare}. We used the BFGS approach when applicable. The CQE indeed recovers the entire spectrum, although there are three states with qualitative errors. The VQE spectra converges, though has large errors above the first four excited states. qEOM \cite{ollitrault2020prra} is based on the VQE-UCCSD ansatz (which is quite accurate for this configuration), and does qualitatively well though the number of states is limited to the dimensionality of the excitation manifold.

\begin{figure}
    \centering
    \includegraphics[scale=0.5]{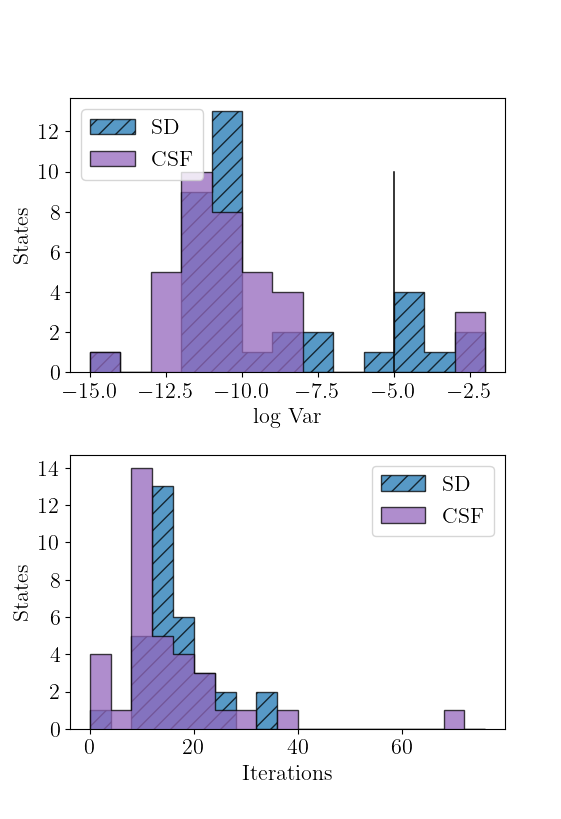}
    \caption{Comparison of Slater determinant and configuration state functions as initial state conditions in rectangular ${\rm H}_4$. (top) Distributions of the variances of the obtained excited states, with the target residual threshold of $10^{-5}$ given by the black bar. (bottom) Comparison of the number of iterations per state. Not included are two SD states with over 150 iterations.}
    \label{fig:3_exact_simulation}
\end{figure}

Within the ES-CQE based approach we found problematic states around $k=16$ to $k=20$ across the dissociation curve with different optimizers. Most of these states were satisfied initially by the ACPSE. While this was expected for the high spin ($S=2$) quintet (which has dimension 1 in this Fock space and is always an eigenstate), this also occurred for other states. To understand this, as well how different initial states affected the accuracy of the procedure, we looked at a spin adapted basis of two-electron operators, as well as the influence of using configuration state functions (SF) or Slater determinants (SD) as initial conditions. Spin adaptation did not address these concerns, and more substantial differences were seen with initial guesses. Fig.~\eqref{fig:3_exact_simulation} shows the accuracy and rate of convergence for all $k-$solutions for different initial states. 
With a residual criteria of $10^{-5}$, the variances show nearly all states are eigenstates. For the CSFs, only three states violated this, and for the Slater determinant basis, eight states were found. Again, it is worth noting that the ACSE and ACPSE are not exact conditions on the SE, in contrast to the CSE and PCSE.  

To understand the violations, we consider the spin and also spatial symmetries of the molecule. The rectangular geometry possesses $D_{2h}$ symmetry. Within the $D_{2h}$ point group, the minimal basis has four different irreducible representations, $A_{1g}$, $B_{1g}$, $B_{2u}$ and $B_{3u}$.  These constrain the state space significantly, which we further discuss in Appendix \ref{sec:symm} and Table~\ref{tab:symm_space}. As a result, there is a subspace of dimension three with symmetries $S=1$, and $\mathcal{R}=A_{1g}$. The ACSE fails within this limited space of three CSFs though not within the entire 6-determinant space. In this instance, the CSFs, while providing a complete set of spin states, do not have sufficient overlap with one another via the two-electron operator, and so spuriously satisfy the ACSE.
\begin{table}[]
    \centering
        \caption{Dimension of the state space for total spin and spatial symmetries. $A_{1g}$ consists of all doubly occupied and the four singly occupied configurations, and the $B$ symmetries belong to all configuration with two singly occupied orbitals.}
    \begin{tabular}{c|ccc}
         & $S=0$ & $S=1$ & $S=2$ \\ \hline
        $A_{1g}$ & 8 & 3 & 1 \\
        $B_{1g}$ & 4 & 4 & 0 \\
        $B_{2u}$ & 4 & 4 & 0 \\
        $B_{3u}$ & 4 & 4 & 0 
    \end{tabular}
    \label{tab:symm_space}
\end{table}
Another challenge is that at least one of the FCI vectors has a clearly problematic structure, especially for a determinant-based approach. The triplet state in question (around $k=15$ across nearly all distances) has the approximate form: 
\begin{equation}
    |\psi_{T} \rangle \approx \frac{1}{\sqrt{2}}(|11000011\rangle - |00111100\rangle ).
\end{equation}
The two main determinants are separated by a quadruple excitation, and cannot be coupled through a single two-body operator. For the SD cases, the large amount of iterations for a few of the runs indicates that the ansatz is not able to easily describe this type of wavefunction. An optimization here requires small back and forth rotations between an intermediate state.

We also checked if a small perturbation around $|\psi\rangle$ could help to move the CSFs from an incorrect solution of the ACPSE. We realized by taking the anti-Hermitian portion of a small two-body matrix and then applying it to the initial state, similar to what would happen in the context of a NISQ device. While the norm of ACPSE could be increased in a controllable manner, the CSF solutions then suffered the same problem as the Slater determinant runs, notably a long convergence. 

\subsection{Choice of Constraint and Convergence Properties}

An important element of the ES-CQE algorithm is the choice of constraint and optimization types. We tested the Lagrangian formulation with constant $\lambda_i$, a penalty based formulation, and an augmented Lagrangian-based approach with the CPSE-based approaches, with results shown in~\ref{tab:constraints_comparison}. We also looked at a deflation based approach within the CQE, and an implementation of the VQD procedure with the UCCSD-ansatz. 

\begin{table}
\caption{Comparison of constraint techniques for rectangular ${\rm H}_4$ with side lengths 1.5 \AA~and 1.0\AA. For different points along a axial stretch, the largest E varies considerably, though for this configuration the energies vary from -2.2 to 0.1 H.  Note the norm of all of the Pauli terms coefficients in $||H||$ was $3.6$, and the Frobenius norm is 56.8.}
\begin{tabular}{cc|cccc}
\multicolumn{2}{c}{Method} & Avg. $\log_{10}$ Var & Avg. Iters. & $k-$ \\
 \hline \hline 
 Lagrangian & $\lambda_i =1$  & -6.1 [1.8] & 9.1 [6.7] & 36 \\
    \hline
  Penalty & $\mu = 1$    & -4.7 [2.9]  & 7.8 [4.1]  & 20 \\
           & $\mu = 4$ & -3.7 [2.5] & 10.7 [6.4] & 36 \\
        & $\mu = 16$ & -4.1 [2.3] & 12.1 [8.0] & 36 \\ 
        & $\mu = 64$& -4.8 [2.4] & 11.6 [8.7] & 27 \\
    \hline
Augmented  & $\mu = 1 $  &     -6.1 [1.9] & 8.7 [5.6] & 36   \\
   Lagrangian      & $\mu = 4$ & -6.3 [1.9] & 11 [7.5] & 36 \\
  ($\lambda_i =1$)      & $\mu = 8 $ & -6.2 [1.9] & 12 [9.1] & 36 \\
    \hline
Deflation & $\beta=1$ & -5.6 [2.2] & 13 [11] & 35 \\
 & $\beta=2$ & -6.1 [1.9] & 11 [11] & 36 \\
       &   $\beta=8$   & -6.5 [1.9] & 17 [15] & 36 \\
      &   $\beta=32$  & -6.7 [2.0] & 26 [30] & 36 \\
      \hline
VQD-UCCSD & $\beta_i  = 4$ & -2.0 [1.0] & 44 [33] & 36 \\
      \hline
\end{tabular}
\label{tab:constraints_comparison}
\end{table}

In general, we found that the constant Lagrangian, augmented Lagrangian, and deflation techniques had the highest accuracy, although differing in their rates of convergence. The penalty-based optimizer, which is commonly used as a heuristic, was problematic even when a large penalty was used, and in some instances was not able to properly converge for each state. One of the advantages of the augmented Lagrangian is that if necessary, the penalty is increased at each iteration until an appropriate solution is found. For the Lagrangian terms, we found $\lambda_i=1$ to be sufficient, even for larger or smaller systems. The deflation based technique was surprisingly robust in finding each solution, although it is clear that the number of iterations depends on the shape of the landscape influenced by the parameter $\beta$.

While these results reflect an average over the entire spectrum of states, we do expect some of the trends to be less pronounced for the first few iterations, when the constraints are not necessarily very strong. For strongly correlated systems however, with high coefficients across several state vectors, we expect that a properly scaled optimization surface is necessary for quick convergence. 
\subsection{Evaluation of $k-$lowest Excited States}

\begin{figure*}
    \centering
    \includegraphics[scale=0.50]{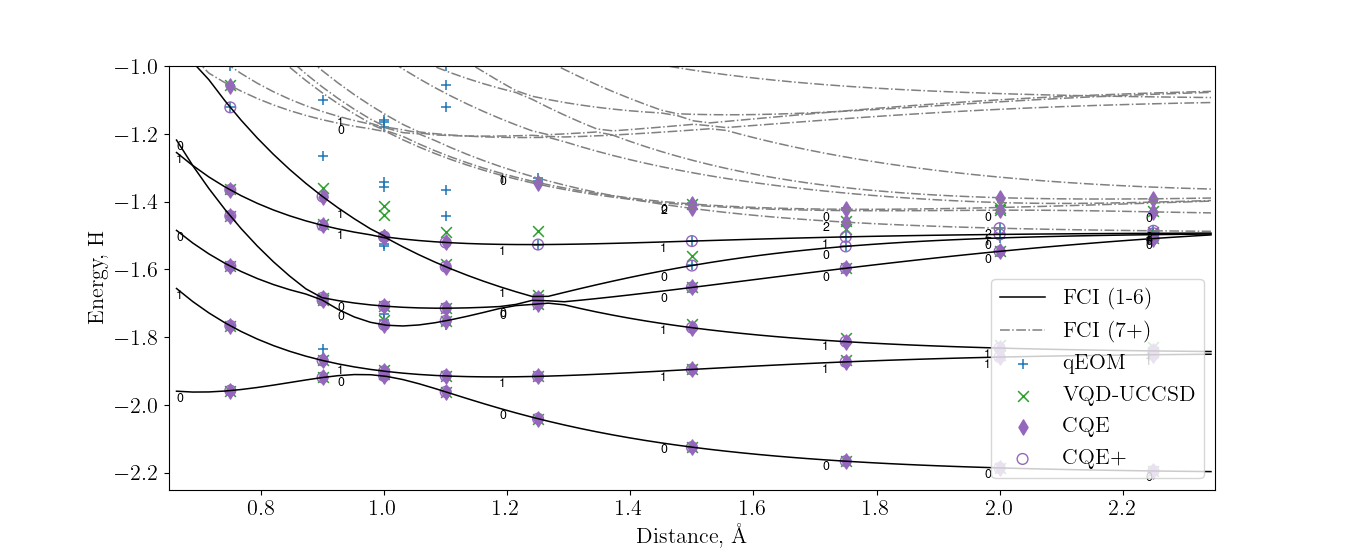}
    \caption{Comparison of $k=6$ excited state calculations for rectangular ${\rm H}_4$, with fixed side length 1\AA. CQE+ denotes the pre-scanned CQE, where twice the number of states are scanned with a low threshold run ($||A||=0.01$, 6 max iterations). The qEOM method was not restricted to dimension 6. The total spin $S$ is indicated by integers slightly below the lines. Gradient thresholds were $10^{-3}$.}
    \label{fig:dissociation}
\end{figure*}

A more common problem is to determine $k-$excited states, and then use the relative energies for further calculations. Continuing our analysis of rectangular ${\rm H}_4$, we evaluate the 6-lowest energy eigenstates in the dimeric dissociation in Figure~\ref{fig:dissociation}. Comparison of the energies are also enumerated in Table~\ref{tab:dissociation}. We evaluated the qEOM, VQD (with a UCCSD ansatz), the ES-CQE (with the augmented Lagragian approach), and the ES-CQE+ (with 6-additional states at $||A|| = 0.01$) methods.

Here, qEOM is not restricted to a manifold of 6-excitations, and so possibly allows for better orderings. Around 1.00\AA, the UCCSD ground state ansatz is not able to achieve millihartree accuracy, and the qEOM gives incorrect results. Additionally a spurious solution to the VQD anastz is found for the next few points. As the energy gaps decrease, the VQD and CQE based approaches begin to converge to higher solutions based on the energy gaps, with qEOM increasing in reliability. With low-depth scans, the CQE+ scheme on average selects much better states, although at the largest separations converges to solutions that are in the range of milli-hartrees higher, instead of tens of milli-hartrees. 

\begin{table}
\caption{Energy comparisons (in millihartrees) of excited state methods with the FCI states for rectangular ${\rm H}_4$. (top) Average energy difference with the $k-$ FCI state. (bottom) Differences with the {\emph closest} FCI energy (no replacement). CQE+ used 12 states with a low threshold run ($||A||=0.01$, 6 max iterations). }
\begin{tabular}{c|cccc}
\hline\hline
\multicolumn{5}{c}{Average $||E_M[k]-E_{\rm FCI}[k]||$, mhartree} \\ 
 \multicolumn{1}{c}{Distance} & qEOM & VQD & CQE & CQE+ \\
 \hline \hline
0.50 & 0.221 & 0.963 & 0.0263 & 0.0263 \\
0.75 & 1.74 & 64.9 & 64.3 & 0.001 \\
0.90 & 1696 & 27.0 & 71.8 & 0.0008 \\
1.00 & 436 & 114 & 20.3 & 0.017 \\
1.10 & 577 & 60.4 & 84.8 & 0.0006 \\
1.25 & 6.560 & 41.171 & 183.1 & 0.0015 \\
1.50 & 3.045 & 197 & 265 & 0.002 \\
1.75 & 2.137 & 307.5 & 190.7 & 0.009 \\
2.00 & 1.579 & 412 & 168 & 21.2 \\
2.25 & 1.074 & 467.4 & 129.3 & 11.4 \\
2.50 & 0.132 & 509 & 56.8 & 9.63 \\
\hline \hline
\multicolumn{5}{c}{Average $\min_j ||E_M[k]-E_{\rm FCI}[j]||$, mhartree} \\
 \multicolumn{1}{c}{Distance} & qEOM & VQD & CQE & CQE+ \\
\hline \hline
0.50 & 0.221 & 0.963 & 0.026 & 0.027 \\
0.75 & 1.742 & 3.198 & 0.000 & 0.001 \\
0.90 & 123 & 24.8 & 0.0005 & 0.0008 \\
1.00 & 230 & 114 & 0.001 & 0.017 \\
1.10 & 130 & 30.5 & 0.0003 & 0.0006 \\
1.25 & 7.13 & 41.2 & 0.0485 & 0.001 \\
1.50 & 3.048 & 30.5 & 0.045 & 0.002 \\
1.75 & 2.137 & 22.994 & 0.0596 & 0.009 \\
2.00 & 1.579 & 11.794  & 0.000 & 0.001 \\
2.25 & 1.0738 & 11.333 & 0.002 & 0.001 \\
2.50 & 0.132 & 12.118 & 0.001 & 0.004 \\
\hline 
\end{tabular}
\label{tab:dissociation}
\end{table}
In Table~\ref{tab:dissociation} we report both the distance to the $k-$lowest eigenstate (upper) as well as the nearest unique eigenstate from the FCI solution (below). The former can be considered as the reliability of the procedure to finding the next lowest energy state, whereas the latter can be considered as the reliability of the solution. In both of these areas we see that the CQE consistently outperforms VQD as well as the qEOM procedures, though qEOM does better at larger distances where more states are captured. The CQE+ itself is a relatively low cost procedure, but improves the selected states substantially. Additionally, the seeded solutions can be used as initial guesses for the default ES-CQE method, likely aiding convergence further. Both the CQE and CQE+ provide highly reliable solutions, even at points where the VQD and qEOM struggle (particularly around 1\AA, where the failure of the qEOM method is significant).  

\section{Conclusions and Outlook}
The current results highlight the accuracy and efficiency of the ES-CQE algorithm, particularly for the rectangular ${\rm H}_4$ system. The current form requires only $O(r^4)$ parameters, which is ideal for scaling to larger systems. Barring highly symmetric cases, the residuals of the ACPSE correspond well with the exact solution, including around strong electron correlation where multi-reference solutions are required. Problematic states do not have sufficient overlap via the set of double excitations, contained in the transition density matrix. In these instances, the state can be invariant to two-body transformations while not being an eigenstate. As mentioned above, practically we can check the variance of the state when the ACPSE is satisfied, or take insights from a CSE-based approach, involving non-unitary manipulations of the states\cite{smart2023}. Another simple approach could involve exploring higher excitations through higher contractions of the CSE (i.e. the 3-5 CSE)\cite{mazziotti1998ijqc}, though at a higher polynomial cost. While the $k-$Brillouion conditions (i.e. anti-Hermitian components of higher order contracted equations)\cite{kutzelnigg1979cpl}, are necessary to describe the complete invariance of a state, in practice $k-$local Hamiltonians are likely dominated by local interactions even for excited states, which we see in the present result. 

These results also have implications for other state-based methods. For instance, the short comings of the residual condition being zero would also be present in the ADAPT-VQE\cite{Grimsley2018} approach, where it is used to select a pool of operators. Such techniques for expanding the excitation pool have similarities to the projected quantum eigensolver approach\cite{stair2021}, where arbitrary excitations can be generated in the operator pool. Within VQE itself, the answer is not so clear, as the VQE gradients do not correspond to the CQE gradients, and one might be able to find the proper rotation given a coupled state. Regarding the projection techniques, the PSE also can be applied to VQE techniques, particularly as it allows for other constrained techniques to be incorporated and does not require any information on the state. We do note though in most cases the deflation technique was fairly accurate, with the main concern being the rate of convergence depending on the constraints. 

The accurate and reliable calculation for ground and excited state solutions for the Schr{\"o}dinger equation is a challenging task, but well suited for explorations on quantum computers. The current approach expands CQE-based approaches, finding excited states by iteratively reducing the optimization space. Using a two-body unitary ansatz, we show that highly accurate excited states can be obtained at a reasonably low cost, involving between $O(r^4)$ and $O(r^6)$ scaling operations, across the entire Hilbert space. We also show several states which violate unitary two-body invariance, necessitating the development of more robust methods for ensuring the accuracy of eigenstates. These states are not common occurences, but might exist in highly symmetric systems. 

Regarding noise and applications for near-term computers, the current work focuses on the limitations of the ideal ES-CQE approach. Understanding the essential limitations and capabilities of the algorithm we wish to run allows us to expand and pursue better near-term, or early fault tolerant applications. For near-term efforts, implementing overlap circuits will be challenging, as these double the effective circuit length, and require additional connectivity via an ancilla qubit. Moving beyond the systems here, we expect this approach to open up other avenues for applications, and look to explore these in quantum many-body systems on NISQ devices. 

\section*{Acknowledgements}
This work is supported by the U.S. Department of Energy Basic Energy Sciences (BES) under grant number DE-SC0019215, and the National Science Foundation RAISE-QAC-QSA under grant number DMR-2037783. 

\appendix
\section{Computational Details}
The CQE and VQD calculations were run using packages based on HQCA (v22.9)\cite{smart}, and Qiskit (v0.22.0) \cite{Qiskit}, with molecular integral generation and FCI references performed with PySCF (v2.0) \cite{Sun2017}. qEOM resulted were also generated with Qiskit. 

\section{Configuration State Functions}
The configuration state functions, which have known ancilla-based efficient implementations \cite{sugisaki2016jpca,sugisaki2019acs}, were implemented using a genealogical coupling scheme \cite{helgaker2000met}. Instead of normal spin tensor operators, which are eigenstates of $\hat{S}^2$, we couple the plus or minus doublet creation operator with the corresponding annihilation operator. That is, using the genealogical coupling scheme, we have:
\begin{equation}
\begin{split}
        \hat{O}^{S,M}_N(t) &=  C^{S,M}_{t_N,\frac{1}{2}}\hat{O}^{S-t_N,M-\frac{1}{2}}_{N-1}(t) \hat{O}^{\frac{1}{2},\frac{1}{2}}_{N} \\
        & + C^{S,M}_{t_N,-\frac{1}{2}}\hat{O}^{S-t_N,M+\frac{1}{2}}_{N-1}(t) \hat{O}^{\frac{1}{2},-\frac{1}{2}}_{N}
\end{split}
\end{equation}
where $C$ are Clebsch-Gordon coefficients. We replace $\hat{O}^{\frac{1}{2},\sigma}_N = a^\dagger_{N\sigma}$ with:
\begin{equation}\label{coupling_op}
    \tilde{O}^{\frac{1}{2},\sigma}_N = a^\dagger_{N\sigma} + a^{}_{N\sigma}
\end{equation}
Though this is not a spin tensor operator anymore, all strings resulting from the coupling procedure with any number of annihilation operators will be zero acting on a closed shell reference, resulting in a CSF. 

To prepare this on a quantum computer, most fermion-to-qubit transformations act on the operator in Eq.~\eqref{coupling_op} to yield a single Pauli string. Each CSF can be described by a vector $t$, which in turn creates a an exponentially scaling number of sequences of $\tilde{O}_N^{v_1} \tilde{O}^{v_2}_{N-1} \cdot \cdot \cdot \tilde{O}^{v_N}_{1}$. Because each $\vec{v}$ is a unique combination of orbitals (i.e. all the created Slater determinants are orthogonal at every step), we can conditionally apply a Pauli string rotation (with the angle yielding the correct coefficients) based on the complete string of the previous sites. As mentioned previously, because the number of recursive pathways scales exponentially with the number of open configurations, this method contains an exponential (w.r.t. the order of the configuration) cost in the preparation. However, this is a one-time cost, and by limiting the preparation to low-spin states, essentially becomes a constant-scaling cost relative to the excited state procedure. 

\section{Symmetries of Rectangular ${\rm H}_4$}\label{sec:symm}
The rectangular configuration belongs to the $D_{2h}$ point group. In STO-3G, the four orbitals have $A_{1g}$, $B_{1g}$, $B_{2u}$ and $B_{3u}$ irreducible representations, which form a group under multiplication. 

For the spin symmetries, we denote the listed orbital occupation of $r$ spatial orbitals with $N$ open shell electrons as the orbital configuration, which generates specific $\hat{S}^2$ and $D_{2h}$ symmetries for each configuration. Table~\ref{tab:orbcon} shows orbital configurations and the irreducible representation of the point group
\begin{table}[]
    \centering
        \caption{Example orbital configurations occupying point groups and their respective irreducible representation for $D_{2h}$ rectangular ${\rm H}_4$ with $\hat{S}_z$=0. The maximum total spin is also given for each configuration. (*) All doubly occupied configurations are $A_{1g}$}

    \begin{tabular}{cccc|cc}    
    \multicolumn{4}{c|}{Occupations (mod 2)} & Irr. & \\
        $A_{1g}$ & $B_{1g}$ &$B_{2u}$ & $B_{3u}$  & Rep. & $S_{max}$ \\ \hline \hline 
       0 & 1 & 1 & 0 &  $B_{3u}$ & 1 \\
       0 & 1 & 0 & 1 & $B_{2u}$ & 1 \\
       0 & 0 & 1 & 1 & $B_{1g}$ & 1 \\
       1 & 0 & 0 & 1 & $B_{3u}$ & 1 \\
       1 & 0 & 1 & 0 & $B_{2u}$ & 1 \\
       1 & 1 & 0 & 0 & $B_{1g}$ & 1 \\
       \hline
       1 & 1 & 1 & 1 & $A_{1g}$ & 2\\
       \hline
       0 & 0 & 0 & 0 &  $A_{1g}$ & 0 \\
       \hline
    \end{tabular}
    \label{tab:orbcon}
\end{table}

The $N=4$ case contains 1 quintet state, 3 triplets, and 2 singlets, with the $A_{1g}$ symmetry. These three triplet states appear to cause most of the problems in the the CSF initial guesses, although for the Slater determinants more states are problematic. We note that to couple two eigenstates via a two-body operator requires that the direct product of irreducible representations of the coupling operator match the product of the two states.  While we can find other $B$ operators to coupled triplet states to the $A_{1g}$ states, the states themselves do not have the proper symmetry, and so are zero in the Hamiltonian.  

To explain the nullity of the three $A_{1g}$ triplet operators, we can express the ACSE condition for a set of states $|p\rangle, |q\rangle$ as:
\begin{align}
    {}^2 A &= \langle p | [\hat{H},\Gam{i}{k}{l}{j}] |p \rangle  \\ &= \sum_{q} H_{pq} \langle q | \Gam{i}{k}{l}{j} | p\rangle - H_{qp} \langle p | \Gam{i}{k}{l}{j} | q \rangle \\
    &= \sum_{q} H_{pq} (\langle q | \Gam{i}{k}{l}{j} | p\rangle - \langle p | \Gam{i}{k}{l}{j} | q \rangle)
\end{align}
where the matrix element $H_{pq}$ is Hermitian. One condition for this to be zero is that the transition matrix for some operator (which is not Hermitian) is Hermitian. If these transition matrices are zero for all operators in the two-electron space, then it is likely that the states have a particular symmetry. CSFs are highly symmetric, and for even number of open-shell electrons have a particle-hole form. 

However, this alone is insufficient, and to the best of our knowledge, only the $S=1$ subspace of this configuration satisfies this condition. We also looked at the 3-TDM with a $N=6$ configuration, and there regularly find non-Hermitian TDMs. Additionally, loosening the $S=1$ constraint to consider all 6 states in the spatial $N=4$ configuration yielded non-Hermitian TDMs. The result indicates that the triplet subspace spuriously fulfills the ACSE, and cannot couple with other triplet states. While this is clearly not a widespread occurrence, it does provide a demonstration of symmetry-based restrictions of the state space. 
\end{document}